\documentclass{article}
\usepackage{spconf,amsmath,graphicx,epsfig,amssymb}
\usepackage{url}

\title{Adaptive Temporal Compressive Sensing for Video}
\name{Xin Yuan, Jianbo Yang, Patrick Llull, Xuejun Liao, Guillermo Sapiro, David J. Brady and Lawrence Carin}
\address{Department of Electrical and Computer Engineering, Duke University, Durham, NC, $27708$, USA}
\begin{document}
%
\maketitle
 \vspace{-0.8cm}
\begin{abstract}
This paper introduces the concept of {\em adaptive temporal compressive sensing (CS)} for video.
We propose a CS algorithm to adapt the compression ratio based on the scene's temporal complexity, computed from the {\em compressed} data, without compromising the quality of the reconstructed video.
The temporal adaptivity is manifested by manipulating the integration time of the camera, opening the possibility to {\em real-time} implementation. The proposed algorithm is a generalized temporal CS approach that can be incorporated with a diverse set of existing hardware systems.
\end{abstract}

\vspace{-0.5cm}
\section{Introduction}
\label{sec:intro}
\begin{figure}[htb]
\centering
  \centerline{\includegraphics[width=8.5cm]{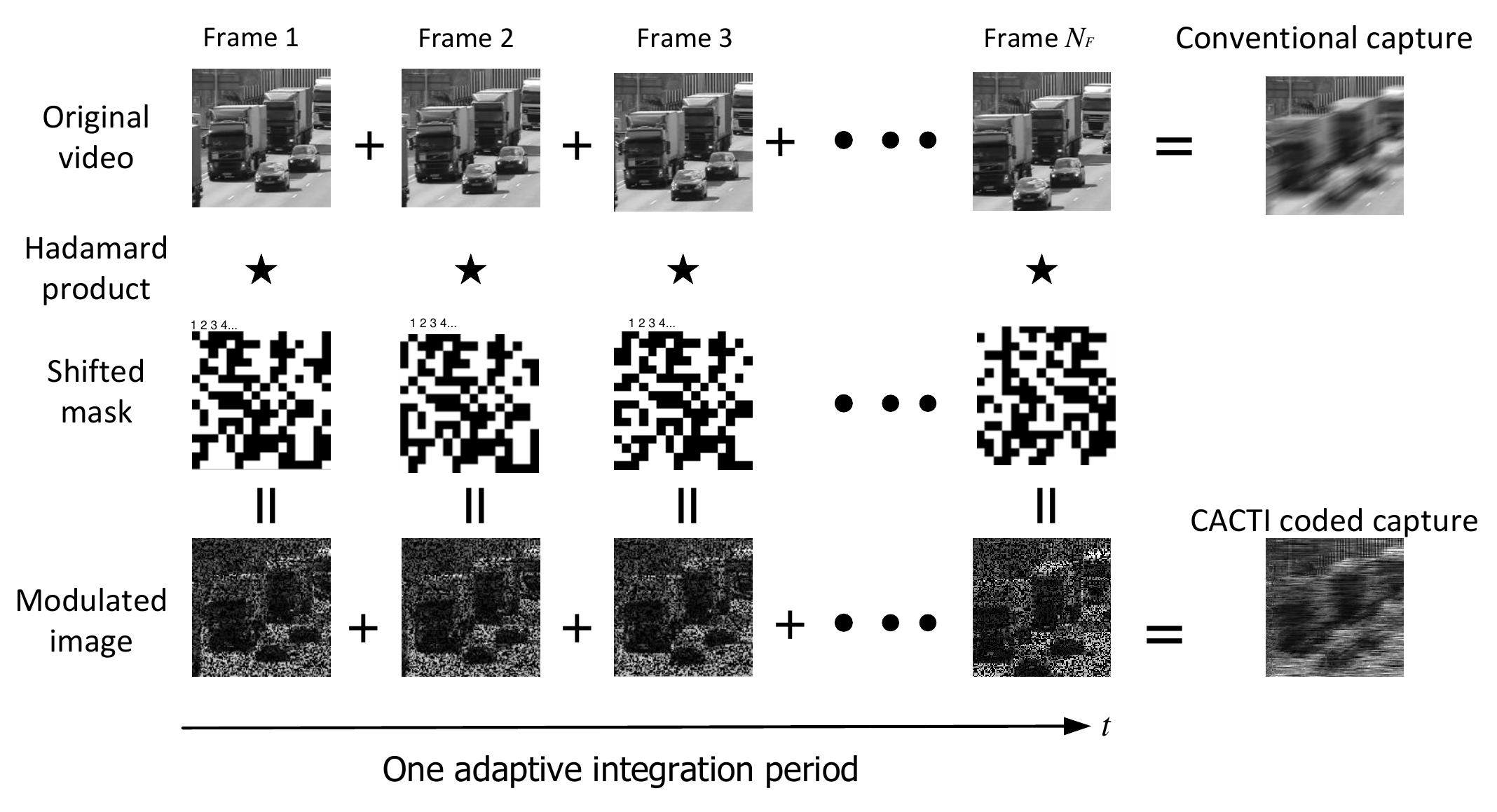}}
  \vspace{-0.3cm}
\caption{\footnotesize{Illustration of the coding mechanisms within the Coded Aperture Compressive Temporal Imaging (CACTI) system \cite{Patrick13OE}. The first row shows $N_F$ high-speed temporal frames of the source datacube video;
the second row depicts the mask with which each frame is multiplied (black is zero, white is one).
In CACTI, the same code (mask) is shifted (from left to right) to constitute a series of frame-dependent codes.
Finally, the CACTI measurement of these $N_F$ frames is the sum of the coded frames, as shown at right-bottom.
}}\medskip
  \label{fig:Dec}
\vspace{-0.5cm}
\end{figure}

\vspace{-0.3cm}
Video compressive sensing (CS), a new application of CS, has recently been investigated to capture {\em high-speed} videos at {\em low frame rate} by means of {\em temporal compression} \cite{Hitomi11ICCV,Reddy11CVPR,Patrick13OE}\footnote{Significant work in spatial compression has been demonstrated with a single-pixel camera \cite{Sankaranarayanan10ECCV,Sankaranarayanan12ICCP,Wakin06PCS,Duarte08SPC}.  Unfortunately, this hardware
cannot decrease the sampling frame rate, and therefore has not been applied in temporal CS.
\cite{Veeraraghavan11TPAMI} achieved compressive temporal superresolution for time-varying periodic scenes by exploiting their Fourier sparsity.}.
A commonality of these video CS systems is the use of {\em per-pixel modulation} during one integration time-period, to overcome the spatio-temporal resolution trade-off in video capture.
As a consequence of active  \cite{Hitomi11ICCV,Reddy11CVPR} and passive pixel-level coding strategies \cite{Patrick13OE,Patrick13COSI} (see Fig.~\ref{fig:Dec}), it is possible to uniquely modulate several temporal frames of a continuous video stream within the timescale of a single integration period of the video camera (using a conventional camera). This permits these novel imaging architectures to maintain high resolution in both the spatial and the temporal domains.
Each low-speed exposure captured by such CS cameras is a linear combination of the underlying coded high-speed video frames.
After acquisition, high-speed videos are reconstructed by various CS inversion algorithms \cite{Tropp07ITT,Bioucas-Dias2007TwIST,liao12GAP,Yang13GMM}.

These hardware systems were originally designed for {\em fixed} temporal compression ratios.
The correlation in time between video frames can vary, depending on the detailed time dependence of the scene being imaged.
For example, a scene monitored by a surveillance camera may have significant temporal variability during the day, but at night there may be extended time windows with no or limited changes.
Therefore, {\em adapting the temporal compression ratio} based on the captured scene is important, not only to maintain a high quality reconstruction, but also to save power, memory, and related resources.

We introduce the concept of {\em adaptive temporal compressive sensing} to manifest a CS video system that adapts to the complexity of the scene under test.
Since each of the aforementioned cameras involves similar integration over a time window, in which $N_F$ high-speed video frames are modulated/coded,
we propose to adapt this time window (the integration time $N_F$), to change the temporal compression ratio as a function of the complexity of the data.
Specifically, we adaptively determine the number of frames $N_F$ collapsed to one measurement, using {\em motion estimation} in the {\em compressed} domain.\footnote{Studies have shown that
improved performance could be achieved when projection matrices are designed to adapt to the underlying
signal of interest \cite{Elad07SPT,Ji08SPT,Carson11SIAM,Duarte-Carvajalino13SPT,Zelnik-Manor11SPT}. However,
none of these methods was developed for video temporal CS.
The adaptive CS ratio for video has been investigated in \cite{Liu11CSVTT,Park09PCS,Azghani10IST,Fowler12FTSP}. Each frame in the video to be captured is partitioned into several blocks based on the estimated motion, and each block is set with a different CS ratio. Though a novel idea, it is difficult to employ it in real cameras since it is hard to sample at different framerates for different regions (blocks) of the scene with an off-the-shelf camera. In contrast, the method presented in this paper can be readily incorporated with various existing hardware systems.}

The algorithm for adaptive temporal CS can be incorporated with a diverse range of existing video CS systems (not only the imaging architectures in \cite{Hitomi11ICCV,Reddy11CVPR,Patrick13OE} but also flutter shutter \cite{raskar2006coded,Tendero13SIAM} cameras), to implement {\em real-time} temporal adaptation.
Furthermore, thanks to the availability of hardware for simple motion estimation \cite{Hsieh92CSVT}, the proposed algorithm can be readily implemented in these cameras.

\vspace{-0.5cm}
\section{Proposed Method}
\vspace{-0.3cm}
\label{Sec:Method}

The underlying principle of the proposed method is to determine the temporal compression ratio $N_F$ based on the motion of the scene being sensed.
In the following, we propose to estimate the motion of the objects within the scene, to adapt the compression ratio for effective video capture.

\begin{figure}[htbp]
\centering
\centerline{\includegraphics[width=8.5cm]{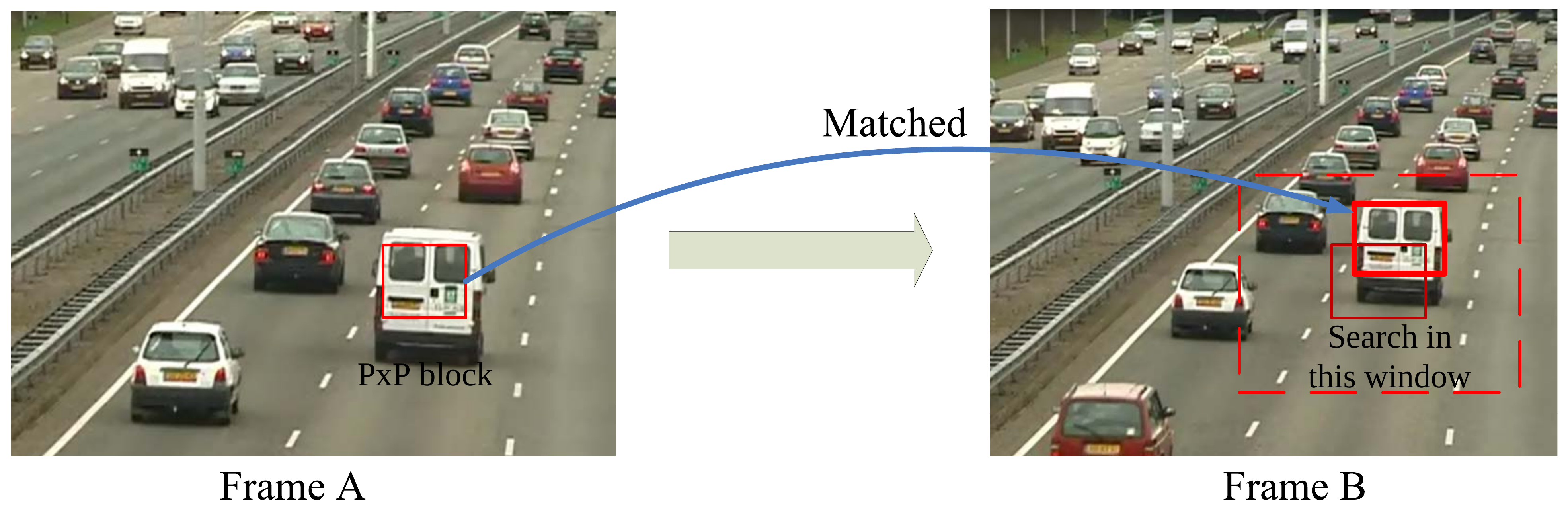}}
 \vspace{-0.3cm}
  \caption{\footnotesize{Basic principle of block-matching.
  Search all the $P\times P$ blocks in the window of frame B to find the one best matched with the block in frame A, and use this to compute the block motion.}}\medskip
  \label{fig:BM}
  \vspace{-0.5cm}
\end{figure}

\vspace{-0.5cm}
\subsection{Block-Matching Motion Estimation}
The block-matching method considered here has been employed in a variety of video codes ranging from MPEG1 / H.261 to MPEG4 / H.263 \cite{Hsieh92CSVT,Ezhilarasan08JCS,Gall92SPIC}.
Diverse algorithms \cite{Ezhilarasan08JCS} have investigated the block-matching concept shown in Fig.~\ref{fig:BM}.
The key steps of the block-matching method are reviewed as follows:
$i$) partition frame A (e.g., previous frame) into $P\times P$ (pixels) blocks;
$ii$) pre-define a window size $M\times M$ (pixels);
$iii$) search all the $P\times P$ blocks in the $M\times M$ windows in frame B (e.g., current frame) around the selected block in frame A;
$iv$) and find the best-matching block in the window according to some metric (e.g. mean squared error), and use this to compute the block motion.
We demonstrate adaptive compression ratios based on this estimated motion from reconstructed video frames in Section \ref{Sec:Result}.

Estimating motion in high-speed dynamic scenes via the block-matching method in the {\em reconstructed} video (after signal recovery) is computationally {\em infeasible} given current reconstruction times at even modest compression ratios.
Hence, we aim to compute the adaption of $N_F$ based {\em directly} on the raw ({\em compressed}) measurements {\em without} the intermediate step of reconstruction.
The following section proposes a method to estimate motion solely on low-framerate, coded measurements from the camera.

\begin{figure}[htbp]
\centering
\centerline{\includegraphics[width=8.5cm]{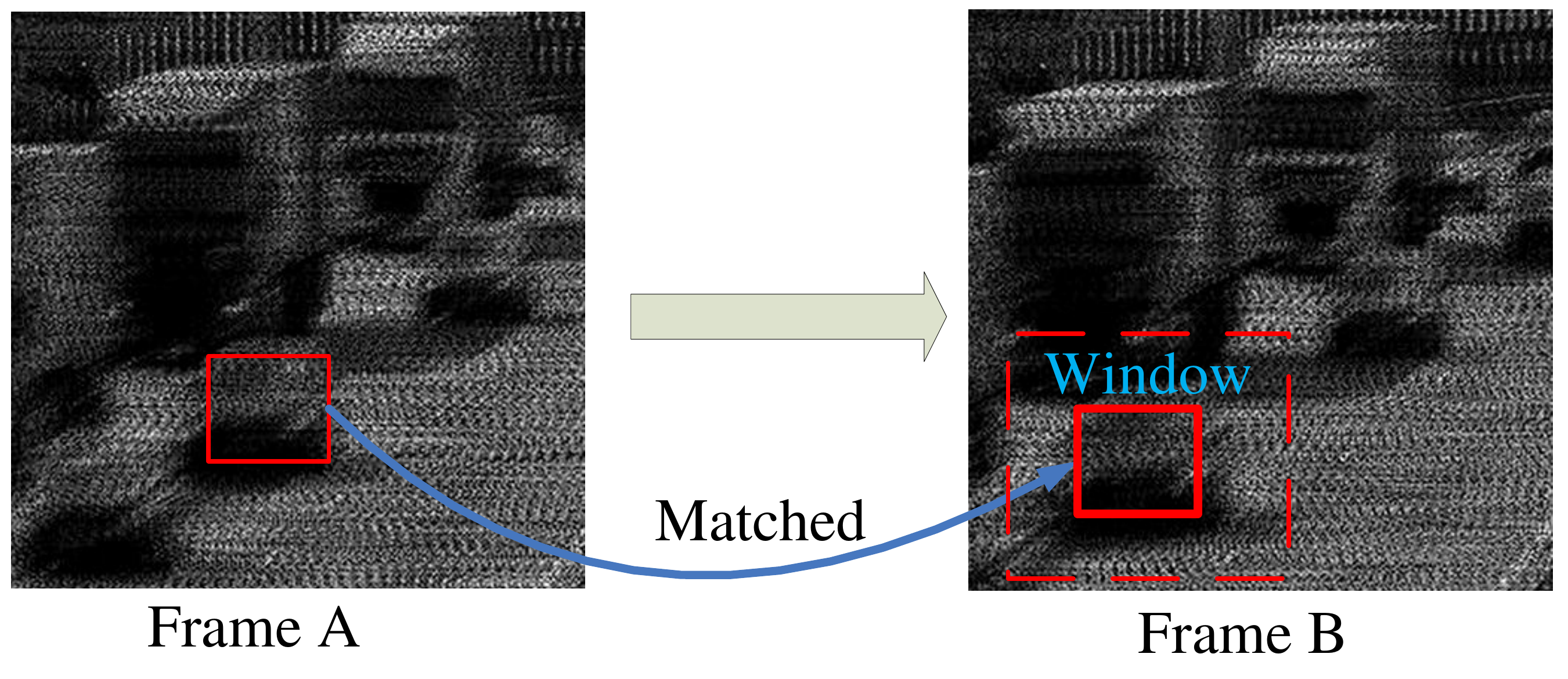}}
\vspace{-0.3cm}
  \caption{\footnotesize{Real-time motion estimation by block-matching.}}\medskip
  \label{fig:BMF}
  \vspace{-0.5cm}
\end{figure}

\vspace{-0.3cm}
\subsection{{\em Real-Time} Block-Matching Motion Estimation}
\vspace{-0.1cm}

Estimating motion from the camera captured data requires the motion to be observable without reconstructing the video frames from the measurement.
Fig.~\ref{fig:BMF} presents the underlying principle of the {\em real-time} block-matching motion estimation approach.
From this figure, it is apparent that the scene's motion is observable within the time-integrated coding structure.
This property lets us employ the block-matching method directly on raw measurements (frames A and B in Fig.~\ref{fig:BMF}) to estimate the scene's motion.
Adapting the compression ratio $N_F$ online is feasible due to the computational simplicity of this method.

\begin{figure}[htbp]
\centering
\centerline{\includegraphics[width=7.5cm]{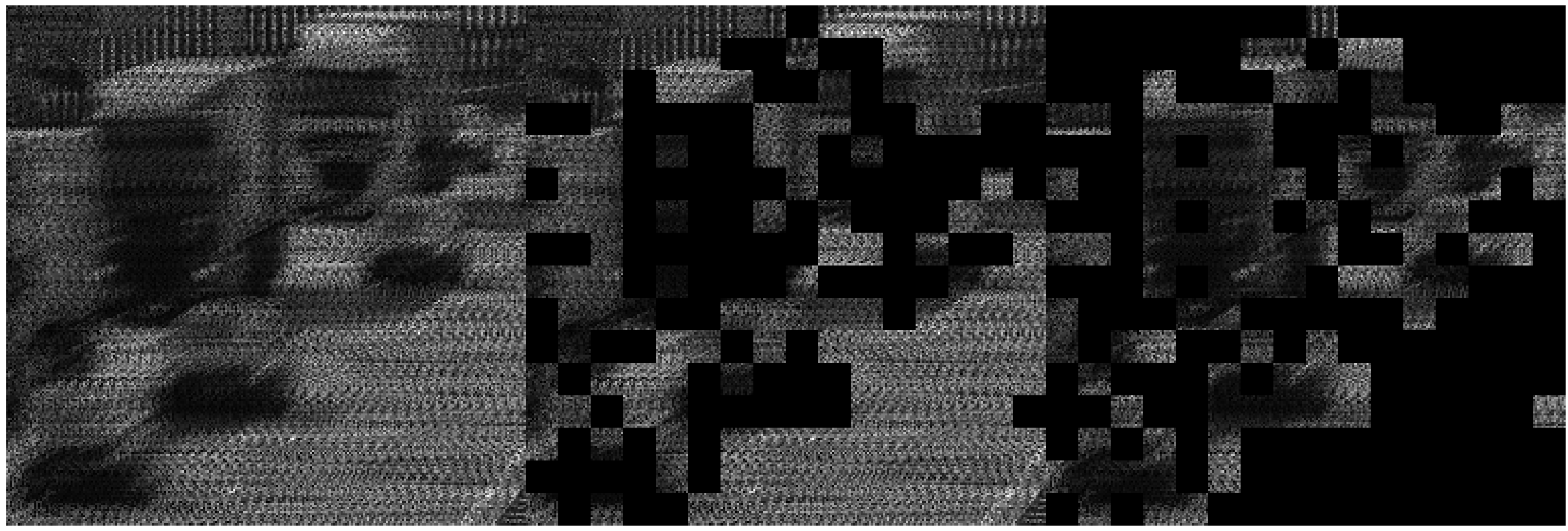}}
  \vspace{-0.3cm}
  \caption{\footnotesize{Segmentation of foreground and background by motion estimation from {\em compressed} measurements. Left is the original measurement; middle is the background blocks with foreground blocks shown in black and the right part presents the foreground blocks with background blocks shown in black. Note that the aim of this work is to estimate the motion, not segmentation. This primary segmentation helps us to localize the moving parts of the scene. $16\times 16$ ($P=16$) blocksize is used and the window size is defined as $40\times 40$ ($M=40$).
  Cross-diamond search algorithm \cite{Cheung02CSVT} has been used to generate this figure and the subsequent results in Section \ref{Sec:Result}.}}\medskip
  \label{fig:Seg_LS}
  \vspace{-0.5cm}
\end{figure}

We may roughly segment the
video sequence into foreground and background regions by computing the
number of pixels that each block traverses between frames.  For each
block, if this number is greater than a pre-specified threshold, the
block is presumably moving and is hence classified as foreground.  Other
blocks are considered background (Fig.~\ref{fig:Seg_LS}).
Notably, we adapt $N_F$ solely based on the estimated motion velocity $V$ (pixels/frame) for the blocks determined by the
block matching algorithm to have moved the greatest number of pixels
between frames.

Intuitively, the compression ratio required to faithfully reconstruct the scene's motion is inversely proportional to the detected velocity $V$, reaching a finite upper bound as $V \rightarrow 0$.
In practice, we simply apply a look-up table to (discretely) appropriately adapt $N_F$ with few computations. See Table \ref{table:Relation} for an example.
Since good hardware exists for motion estimation~\cite{Hsieh92CSVT} , the proposed method can be implemented in {\em real time}.

\begin{figure}[htbp!]
\centering
\centerline{\includegraphics[width=7.5cm]{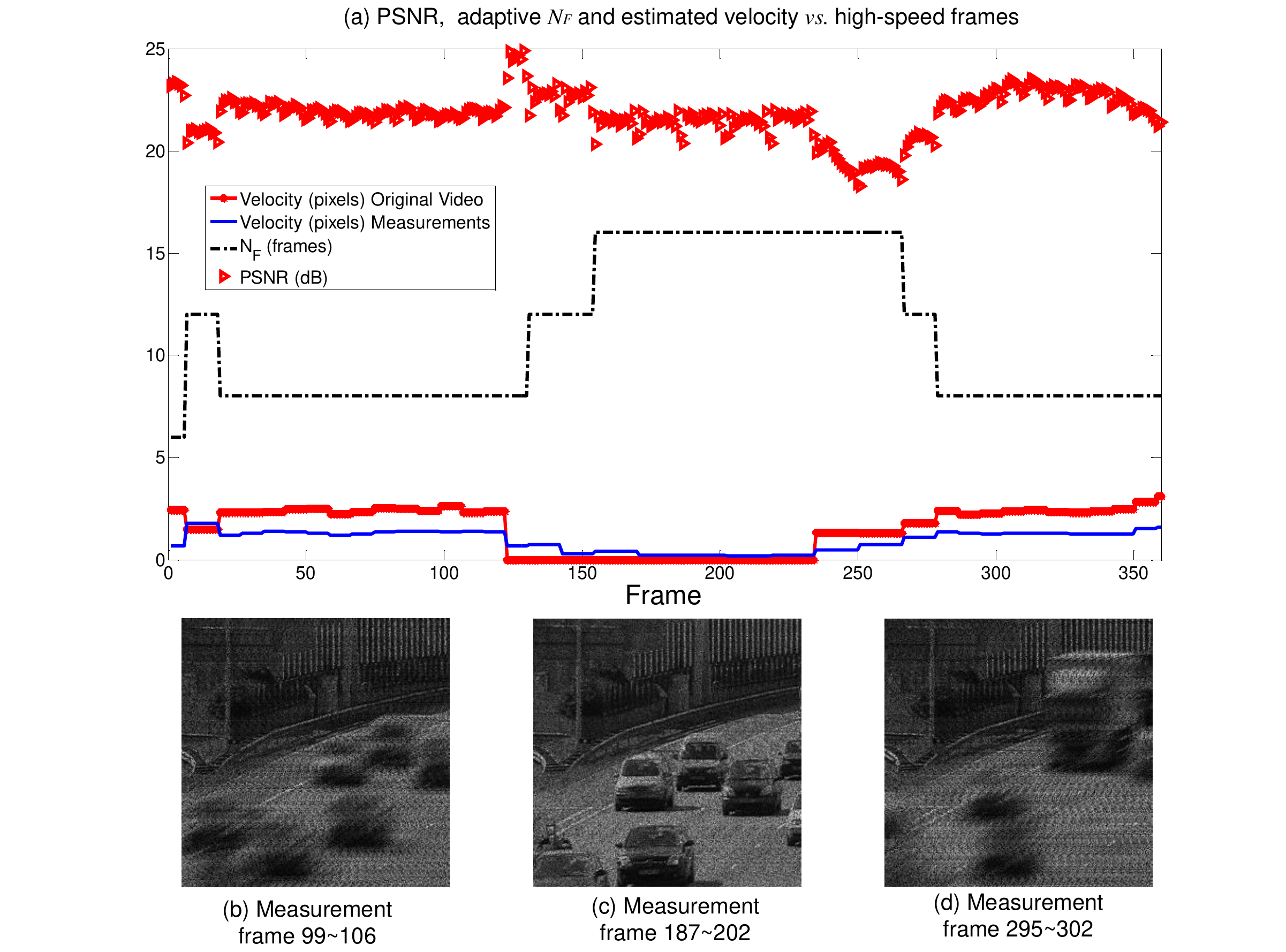}}
  \vspace{-0.3cm}
  \caption{\footnotesize{(a) Reconstruction PSNR (dB), adaptive $N_F$ (frames), and velocities (pixels/frame) estimated from the original video and {\em measurements}, all are plotted against frame number.
  (b-d) Measurements with vehicles at different velocities.}}\medskip
  \label{fig:AdaMea}
  \vspace{-0.5cm}
\end{figure}

It is worth noting that the estimated motion (and hence the compression ratio determined by the look-up table) are used in the upcoming frames.
We assume the consistent motion of the adjacent frames in the video. Sudden changes of the motion will result in one integration time delay of the $N_F$ adaption.
Simulation results in Fig.~\ref{fig:AdaMea} verify this point.
We can of course put an upper bound in $N_F$.

\begin{figure}[htbp!]
\centering
\centerline{\includegraphics[width=7.5cm]{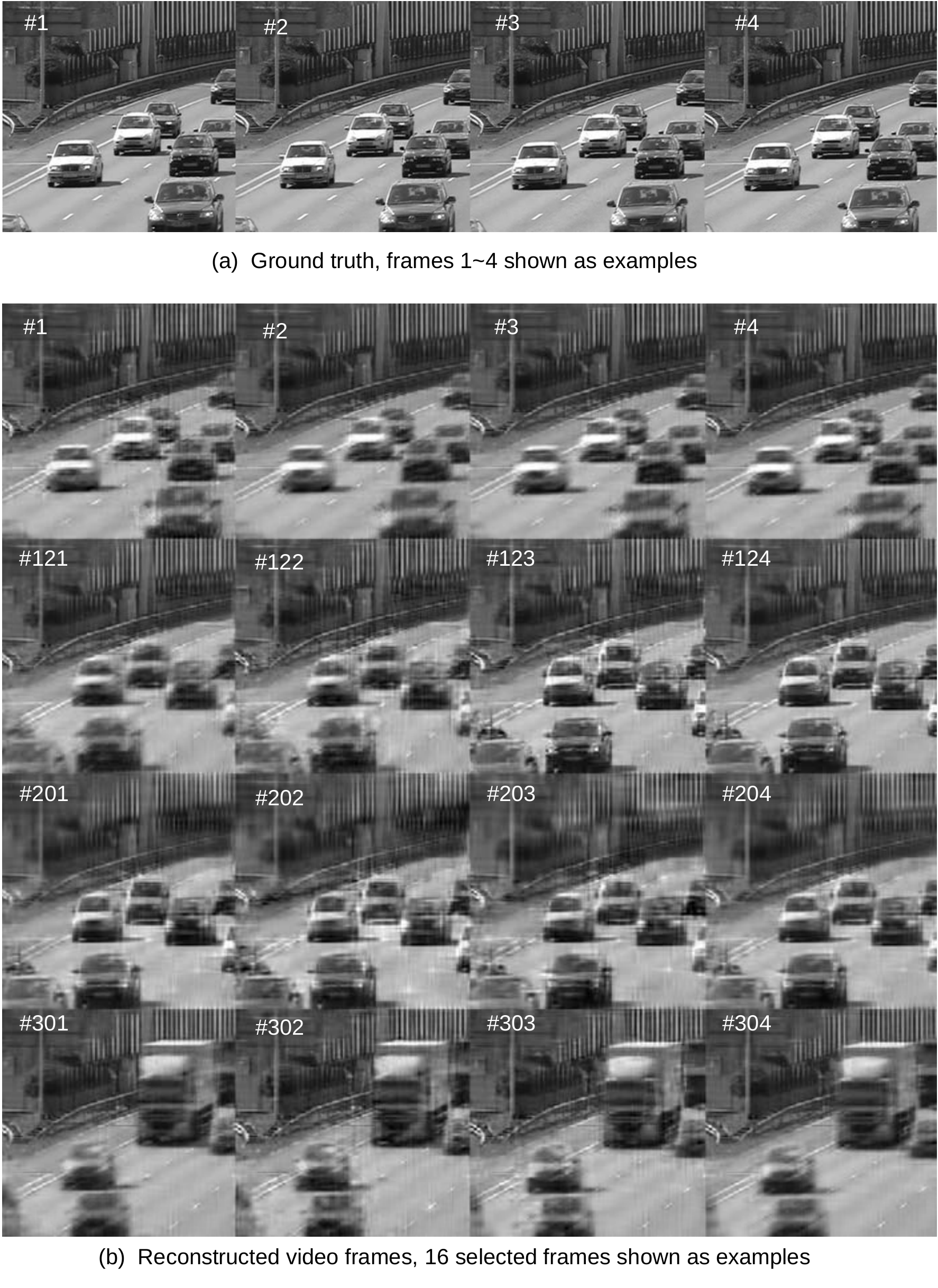}}
  \vspace{-0.3cm}
  \caption{\footnotesize{Selected reconstructed frames (b) based on the adaptive $N_F$ presented in Fig.~\ref{fig:AdaMea}. Frames 1 to 4 in (a) are shown as examples of ground truth.}}\medskip
  \label{fig:Frames}
  \vspace{-0.5cm}
\end{figure}

\begin{figure}[htbp]
\centering
\centerline{\includegraphics[width=7.5cm]{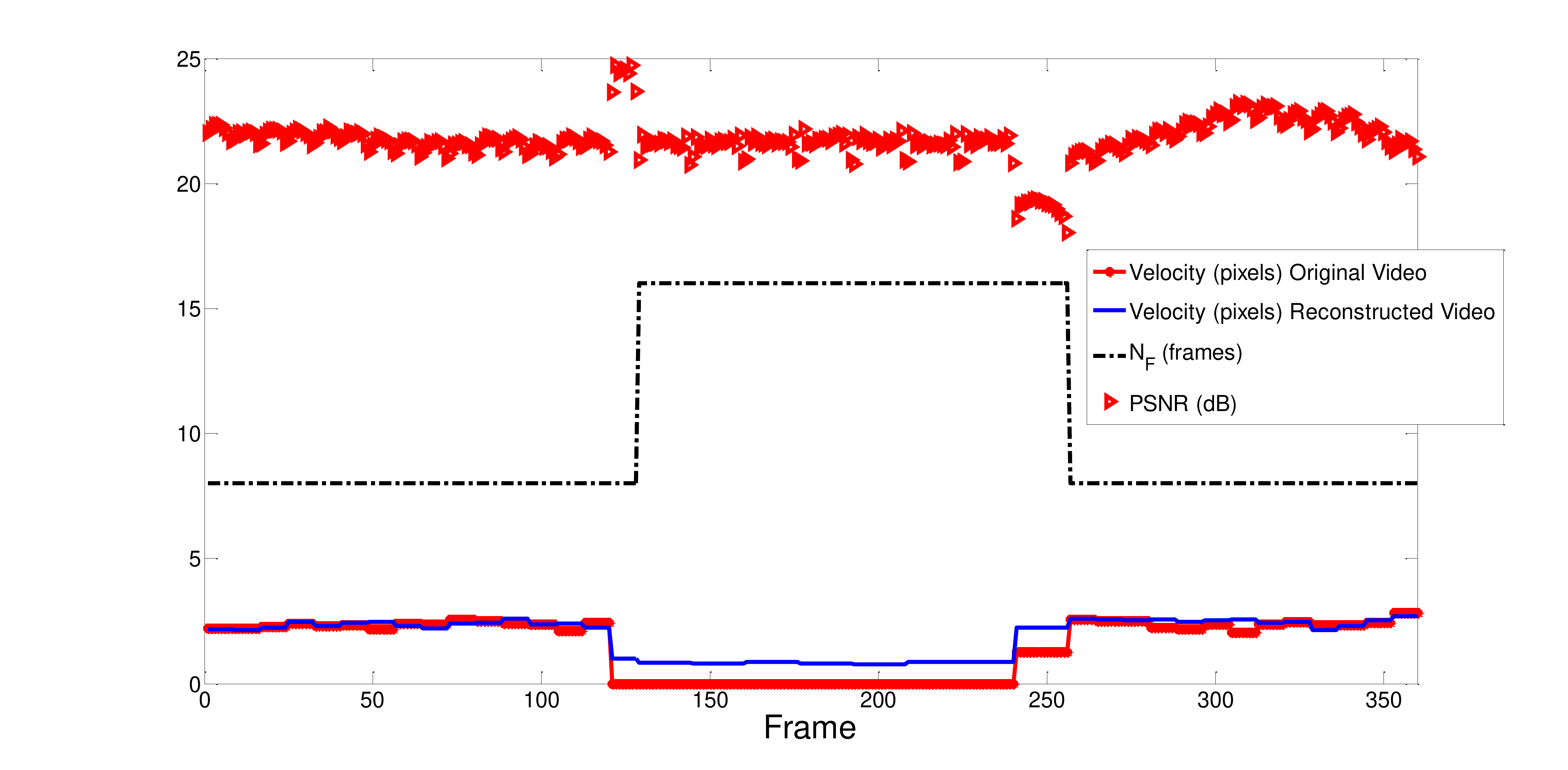}}
    \caption{\footnotesize{Reconstruction PSNR (dB), adaptive $N_F$ (frames), and velocities (pixels/frame) estimated from the original and {\em reconstructed} video frames, all are plotted against frame number.}}\medskip
  \label{fig:AdaRecon}
  \vspace{-0.5cm}
\end{figure}

\begin{table}[htbp!]
\centering
\begin{tabular}{|c|c|c|c|c|c|c|}
\hline
$V$  & $[0,0.5)$ & $[0.5,1)$ & $[1,2)$ & $[2,3)$ & $[3,7)$ & $\ge$ 7 \\
\hline
$N_F$  & 16 & 12 & 8 & 6 & 4 & 2 \\
\hline
\end{tabular}
\vspace{-0.2cm}
\caption{\footnotesize{Relationships between the velocity $V$ (pixels/frame) of the foreground and the compression ratio $N_F$ (frames).}}
\label{table:Relation}
\vspace{-0.5cm}
\end{table}

\begin{figure}[htbp!]
\centering
\centerline{\includegraphics[width=9.5cm]{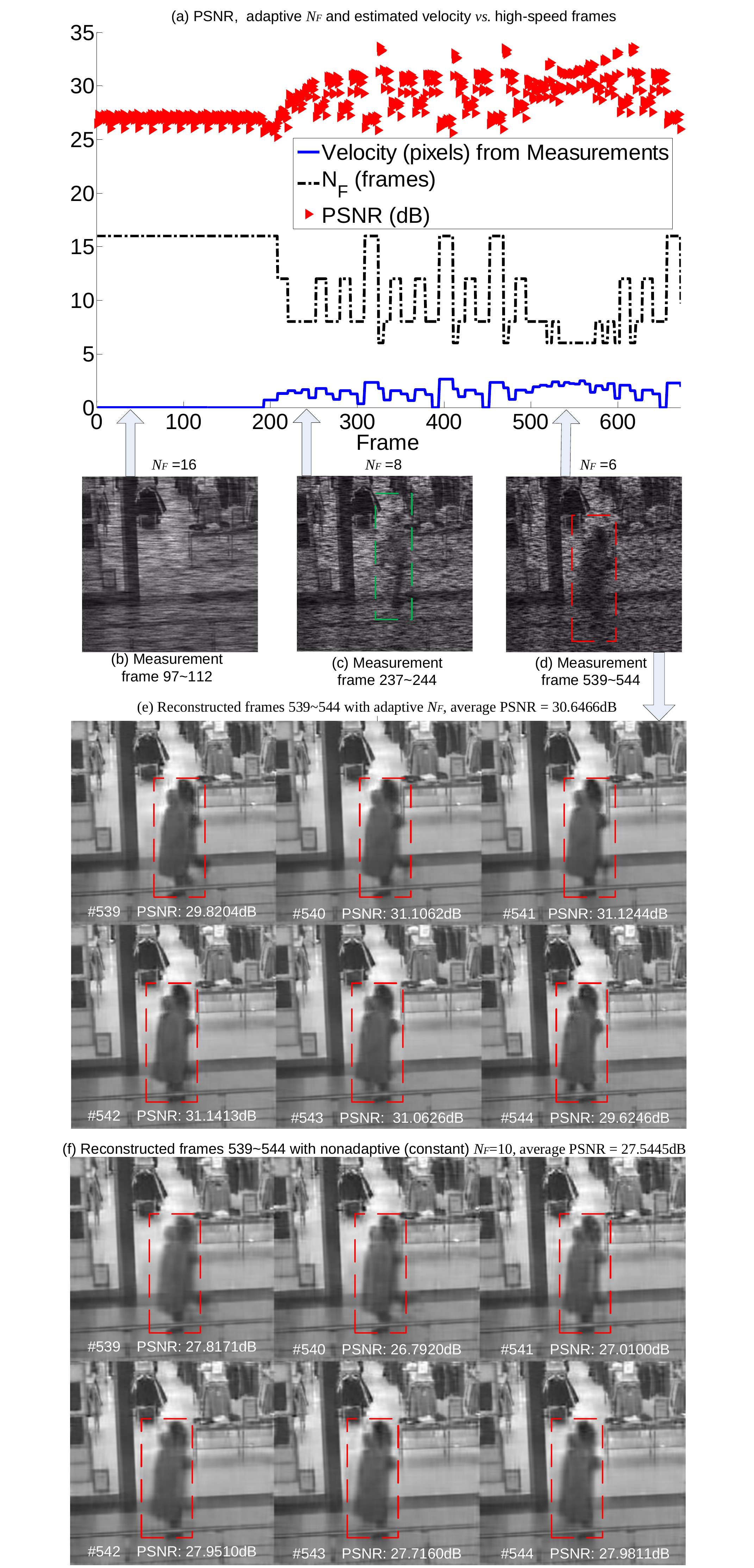}}
   \caption{\footnotesize{Motion estimation and adaptive $N_F$ from the measurements. (a) Reconstruction PSNR (dB), adaptive $N_F$ (frames) (average adaptive $N_F$=10.12), and velocities (pixels/frame) estimated from the {\em measurements}, all are plotted against frame number. (b-d) Measurements when there is nothing, one person, and a couple moving inside the scene, adapted $N_F=16,8,6$, respectively. (e) Reconstructed frames 539-544 from the measurement in (d) with {\em adaptive} $N_F$. (f) Reconstructed frames 539-544 with {\em nonadpative} (constant) $N_F=10$. }}
  \label{fig:AdaMeaShop}
\end{figure}

\vspace{-0.3cm}
\section{Experimental Results}
\label{Sec:Result}
\vspace{-0.2cm}
From \cite{Patrick13OE}, we have found that (based on extensive simulations) shifting a fixed mask is as good as using the more sophisticated time evolving codes used in \cite{Hitomi11ICCV,Reddy11CVPR}.
For convenience (but not necessity), the subsequent results will use a shifted mask to modulate the high-speed video frames. 
\vspace{-0.3cm}
\subsection{Example 1: Synthetic Traffic Video}
\vspace{-0.2cm}
We illustrate the adaptive compression ratio framework on a traffic video \cite{WebVideo} that has 360 frames.
We artificially vary the foreground velocity for this video to evaluate the proposed method's performance for motion estimation and $N_F$ adaption.
Frames 1-120 (Fig.~\ref{fig:AdaMea}(b)) and 241-336 (Fig.~\ref{fig:AdaMea}(d)) run
at the originally-captured framerate; we freeze the scene between frames 121-240 (Fig.~\ref{fig:AdaMea}(c)).
The generalized alternating projection (GAP) algorithm \cite{liao12GAP} is used for the reconstructions.

Table~\ref{table:Relation} provides the compression ratio $N_F$ corresponding to several scene velocities $V$.
This look-up table (learned based on training data\footnote{We use other traffic videos playing at different velocities (different framerates) to learn this table. The main steps are presented as follows: $i$) generate  videos with different motion velocities by changing the framerate;
$ii$) estimate the motion velocities $V$ of the generated videos;
$iii$) modulate the generated videos with shifting masks and constitute measurements with diverse $N_F$;
$iv$) reconstruct the videos with GAP \cite{liao12GAP} from these compressed measurements and calculate the PSNR of the reconstructed video;
$v$) and build the relations between estimated velocities $V$ and $N_F$ maintaining a constant PSNR (around 22dB).
}) seeks to maintain a constant reconstruction peak signal-to-noise ratio (PSNR) of 22dB.
Fig.~\ref{fig:AdaMea} presents the {\em real-time} motion estimation results using simulated low-framerate coded exposures of the traffic video with an initial compression ratio $N_F=6$. 
After a short fluctuation, the estimated velocity of the scene becomes constant; $N_F$ accordingly stabilizes at 8.
When the vehicles freeze, the block-matching algorithm senses zero change in the pixel position and updates $N_F$ to 16.
$N_F$ returns to 8 upon continuing video playback at normal speed.
We can also observe the consistence of velocities estimated from the original video and from the compressed measurements in Fig.~\ref{fig:AdaMea}(a).
Sudden changes in the video's framerate (and hence the motion velocity $V$) are reflected in short fluctuations of the PSNR (for one time-integration period) in Fig.~\ref{fig:AdaMea}(a).
The average PSNR of the reconstructed frames in Fig.~\ref{fig:AdaMea} is 21.8dB, very close to our expectation (22dB).
Fig.~\ref{fig:Frames} presents several reconstructed frames based on the adaptive $N_F$ in Fig.~\ref{fig:AdaMea}.

We additionally evaluate the block-matching algorithm's performance by deploying it to reconstructed frames.
Fig.~\ref{fig:AdaRecon} demonstrates that its performance is similar to the phenomena shown in Fig.~\ref{fig:AdaMea}.
This justifies that it is unnecessary to reconstruct each measurement prior to updating $N_F$.

\vspace{-0.4cm}
\subsection{Example 2: Realistic Surveillance Video}
\vspace{-0.2cm}
Fig.~\ref{fig:AdaMeaShop} implements adaptive $N_F$ on video data captured in front of a shop \cite{WebVideo2}.
Table~\ref{table:Relation} is again useful for this example.

The first 189 frames of this video (Fig.~\ref{fig:AdaMeaShop}(b)) are stationary; nothing is moving within the scene.
As seen before, since $V = 0$, the compression ratio remains at $N_F$=16.
After the 189$^{th}$ frame, different people begin to walk in and out of the video area (Fig.~\ref{fig:AdaMeaShop}(c-d)).
The compression ratio $N_F$ is adapted between 6 and 16 according to the estimated velocity.
When one person walks into the shop (Fig.~\ref{fig:AdaMeaShop}(c)), the compression ratio drops ($N_F$=8).  This results in a better-posed reconstruction of the underlying video frames.  When a couple walks in front of the shop (Fig.~\ref{fig:AdaMeaShop}(d)), $N_F$ drops further to 6. The corresponding measurement and reconstructed frames are shown in Fig.~\ref{fig:AdaMeaShop}(d,e).

This video takes a total of 67 adaptive measurements to capture and reconstruct 678 high-speed video frames, achieving a mean compression ratio $N_F\approx$10.12.
To demonstrate the utility of adapting $N_F$ based on the sensed data, we compare adaptive reconstructions to those obtained when $N_F$ is {\em fixed} at or near its expected value. Fig.~\ref{fig:AdaMeaShop}(f) shows reconstructed frames 539-544 when fixing $N_F$ = 10. Comparing part (e) with part (f), we notice that adapting $N_F$ provides a (3dB) higher reconstruction quality (average PSNR=30.65dB) than fixing $N_F$ near its expected value (average PSNR=27.54dB).
These improvements are most noticeable whenever there is motion within the scene
and demonstrate the potency of temporal compression ratio adaptation in realistic applications.


\vspace{-0.3cm}
\section{Conclusion}
\vspace{-0.3cm}
We have introduced the concept of adaptive temporal compressive sensing for video and demonstrated a {\em real-time} method to adjust the temporal compression ratio for video compressive sensing.  By estimating the motion of objects within the scene, we determine how many measurements are necessary to ensure a reasonably well-conditioned estimation of high-speed motion from lower-framerate measurements.

A block-matching algorithm estimates the scene's motion directly from the compressed measurements to obviate real-time reconstruction, thereby significantly reducing the expended real-time computational resources.
Simulation results have verified the efficacy of the proposed adaption algorithm.
Future work will seek to embed this real-time framework into the hardware prototype.

\bibliographystyle{IEEEbib}
\small

\end{document}